\documentclass{aip-cp}

\usepackage[numbers]{natbib}
\usepackage{rotating}
\usepackage{graphicx}

\begin{document}

\title{Charge and spin asymmetries in elastic lepton-nucleon scattering}

\author[aff1]{Oleksandr Koshchii\corref{cor1}}
\author[aff1]{Andrei Afanasev}

\affil[aff1]{The George Washington University, Washington, D.C. 20052, USA}
\corresp[cor1]{Corresponding author: koshchii@gwmail.gwu.edu}

\maketitle

\begin{abstract}
Elastic lepton scattering off of a nucleon has proved to be an efficient tool to study the structure of the hadron. Modern cross section and asymmetry measurements at Jefferson Lab require effects beyond the leading order Born approximation to be taken into account. Availability of unpolarized beams of both electrons and positrons in respective experiments would enable to reduce systematic uncertainties due to higher-order charge-odd contributions. In addition, information on an unpolarized electron-to-positron cross section ratio could serve as a test for theoretical models that provide predictions for charge-dependent radiative corrections to elastic lepton-nucleon scattering. Availability of polarized beams of leptons would allow for even more comprehensive study of higher-order effects as some of them are dominant in polarized lepton-nucleon scattering asymmetries. We present a brief overview of effects due to the lepton's charge and target's polarization on elastic lepton-nucleon scattering measurements.
\end{abstract}

\section{INTRODUCTION}

Generally, most of unpolarized elastic electron/positron scattering measurements are analyzed in a framework of the one-photon exchange (OPE) approximation, shown in Fig. 1(a). This means that, in order to extract an equivalent OPE form, one needs to apply radiative corrections to the measured cross sections. These corrections include contributions that are generated by exchanges of virtual particles (Fig. 1(f)-(k)), as well as by unavoidable background contributions coming from the emission of real photons, called bremsstrahlung (Fig. 1(b)-(e)). Due to the infrared-divergent nature of radiative corrections, it is common to separate ``soft'' and ``hard'' photon radiation events. It should be mentioned here that there is no unique approach in separating the photon phase space into soft and hard regions - the most common prescriptions are those of Tsai \cite{TsaiRadCor1961} and Maximon and Tjon \cite{MaximonRadCor2000}.
\begin{figure}[h]
 \centerline{\includegraphics[width=410pt]{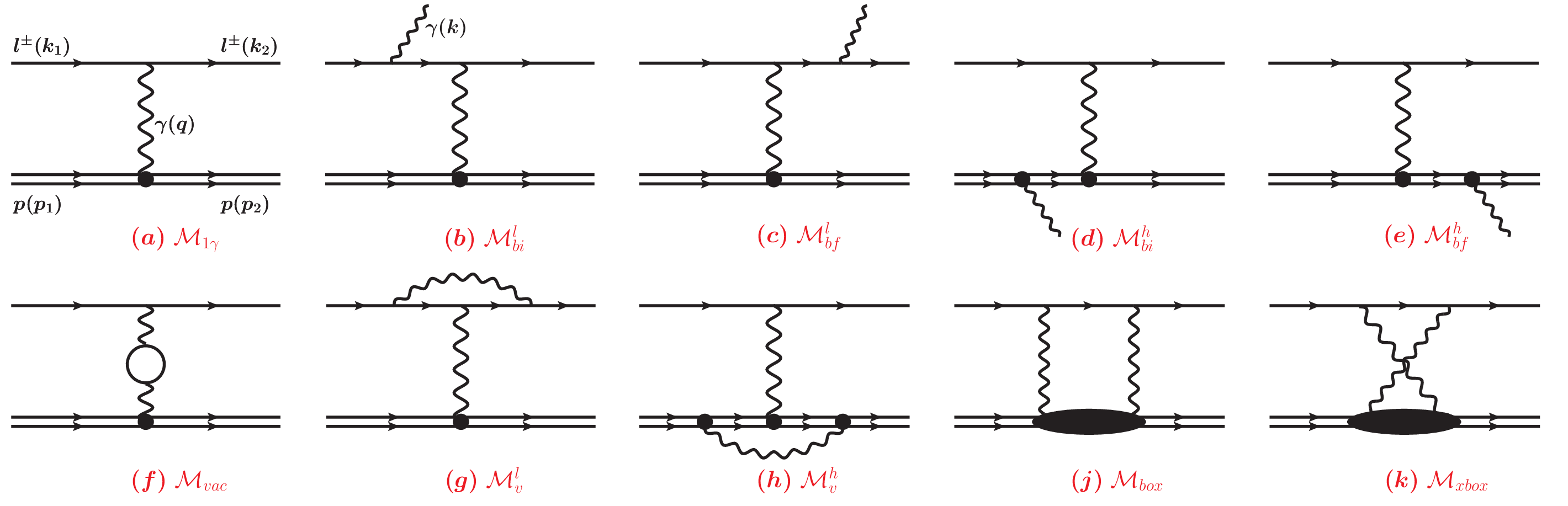}}
 \caption{Leading and next-to-leading-order QED Feynman diagrams describing elastic lepton-nucleon scattering: (a) One-photon exchange, (b)-(c) Lepton bremsstrahlung, (d)-(e) Nucleon bremsstrahlung, (f) Vacuum polarization, (g) Lepton vertex correction, (h) Proton vertex correction,  (j)-(k) Two-photon exchange.}
\end{figure}

The soft-photon contributions are infrared-divergent and independent of the structure of the hadron. They are generally well-understood and can be calculated analytically. In contrast, the hard-photon contributions are finite and nucleon structure dependent. In practice of data analysis, the hard photon radiation effects can be minimized by experimental methods. However, due to finite detector resolution, a complete removal of such effects by pure experimental methods is not possible. In addition, accounting for a realistic detector geometry requires a complicated integration over the phase space of the emitted hard photon. As a result, we are constrained and must use the Monte Carlo (MC) technique to deal with this problem (see, e.g., Ref. \cite{AkushevichElradgen2012, GramolinGenerator2014, AkushevichPRad2015}).

Recently, in unpolarized electron-proton scattering, a lot of attention \cite{CarlsonTPE2007, ArringtonTPE2011, AfanasevTPE2017, RachekTPE2015, RimalTPE2017, HendersonTPEOlympus2017} has been brought to the two-photon exchange (TPE) corrections (Fig. 1(j)-(k)) beyond the soft-photon approximation contributions, which are usually incorporated in standard radiative corrections. These hard TPE contributions, which can provide a percent-level correction to the Born cross section, are believed to affect significantly the discrepancy in proton's electric-to-magnetic form factor ratio \cite{BlundenTPE2003, ChenTPE2004, AfanasevTPE2005} and possess a proper magnitude to be included in future precision measurements. Despite substantial theoretical efforts being directed at understanding of the physics of TPE there is currently no complete calculation valid at all kinematics. More investigation on both theoretical and experimental forefront is needed.

Besides affecting significantly unpolarized lepton-proton scattering, TPE plays an important role in polarized scattering measurements. As it was pointed out by de Rujula et al. about three decades ago in Ref. \cite{RujulaSSA1971}, the imaginary part of the TPE amplitude dominates single-spin asymmetry (SSA) observables when either the beam or target is polarized perpendicularly to the lepton scattering plane in respective elastic scattering. This property of transverse SSAs opens up a unique opportunity to study TPE, as well as provides a direct access to effects beyond TPE, given that the beams of both polarities are available.

\section{Lepton mass effects in unpolarized lepton-proton scattering}

The square of the amplitude that describes elastic lepton-proton scattering and includes all the leading order radiative corrections can schematically be written as
\begin{equation}\label{1}
\big| M \big|^2 = \big| M_{1 \gamma} \big|^2 + \big| M_{b}^l \big|^2 + \big| M_{b}^h \big|^2 + 2 \mathrm{Re} \big[ M_{1 \gamma}^\dag M_{vac} \big] + 2 \mathrm{Re} \big[ M_{1 \gamma}^\dag M_{v}^l \big] + 2 \mathrm{Re} \big[ M_{1 \gamma}^\dag M_{v}^h \big] + 2 \mathrm{Re} \big[ M_{1 \gamma}^\dag M_{2 \gamma} \big] + 2 \mathrm{Re} \big[ (M_{b}^l)^\dag M_{b}^h \big].
\end{equation}
It appears that among all the summands in Eq. (\ref{1}) only the interference between OPE and TPE amplitudes and between lepton and proton bremsstrahlung radiation (the last two terms) are the only charge-odd contributions. This means that hard TPE corrections can be directly extracted by studying the charge asymmetry between elastic $l^+ p$ and $l^- p$ scattering cross sections after respective radiative corrections are applied. Until recently, radiative corrections calculated according to \cite{TsaiRadCor1961} or \cite{MaximonRadCor2000}, which both assume the ultra-relativistic approximation (the lepton's mass $m$ is much smaller than its energy $\varepsilon_1$), were sufficient enough. However, due to advances in technologies and increasing precision requirements of modern experiments, there is an ongoing demand in updating existing MC codes with contributions that do not employ the $m \ll \varepsilon_1$ approximation. In the following subsections we present the results of our update to a MC generator called Elradgen 2.0 \cite{AkushevichElradgen2012} that accounts for the mass of the lepton in elastic $l^\pm p$ scattering. The update is essential for the future MUSE experiment \cite{GilmanMUSE2013} that is going to measure simultaneously elastic $e^\pm p$ and $\mu^\pm p$ scattering.

\subsection{Soft TPE and bremsstrahlung}

In the soft-photon exchange approximation, both charge-odd terms in Eq. (\ref{1}) can be factorized by the square of the one-photon exchange amplitude, which means that we can write down the charge-dependent cross sections as
\begin{equation}\label{2}
d \sigma^\pm = d \sigma_{1 \gamma} (1 \pm \delta_{ch}),
\end{equation}
where the charge-odd contribution is given by the respective asymmetry
\begin{equation}\label{3}
\delta_{ch} = \frac{d \sigma^+ - d \sigma^-}{d \sigma^+ + d \sigma^-}.
\end{equation}
We have accounted for the influence of the lepton's mass on model-independent contributions to the charge asymmetry $\delta_{ch}$ in unpolarized $l^\pm p$ scattering \cite{KoshchiiAsymmetry2017}. The calculation was performed according to the approach of Tsai. The graphical representation of our results for the asymmetry in kinematics of MUSE is shown in Fig. 2(c). Moreover, our MC generator had been re-adjusted to account for the lepton mass effects in charge-even contributions that consist of vacuum polarization, lepton vertex, and lepton bremsstrahlung corrections. Respective radiative corrections, which also include charge-odd contributions \cite{KoshchiiAsymmetry2017}, are shown in Fig. 2(a)-(b).
\begin{figure}[h]
 \centerline{\includegraphics[width=350pt]{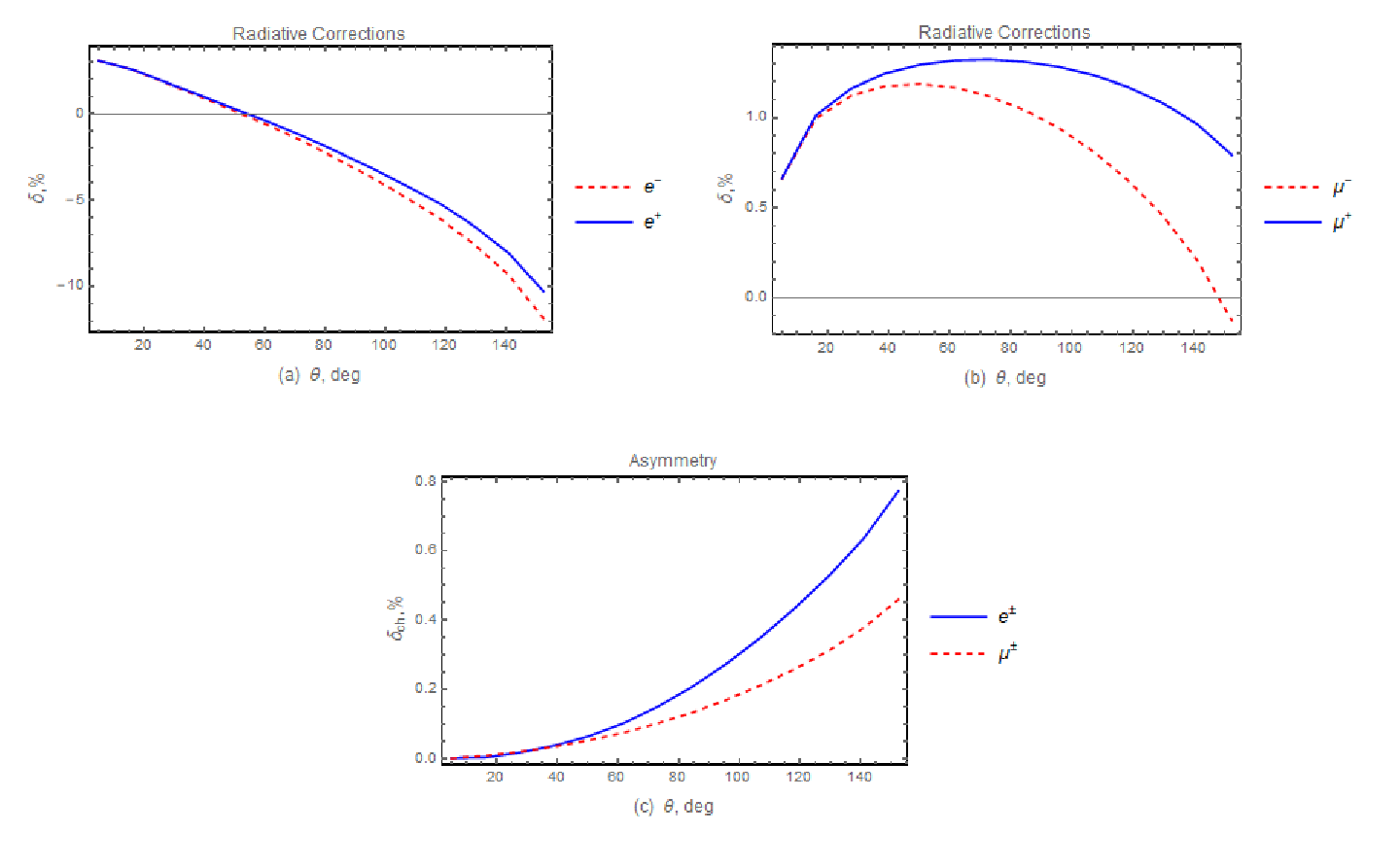}}
 \caption{Radiative corrections and the soft-photon approximation asymmetry in unpolarized $l^\pm$ scattering. The beam's momentum is $|\vec{k}_1| = 115$ MeV.}
\end{figure}

It can clearly be seen from Fig. 2(c) that the lepton mass effects cannot be neglected in the considered kinematical setting whenever the precision goal is set on a sub-percent level. Moreover, the results shown in Fig. 2(a)-(b) demonstrate that the non-zero mass of the lepton considerably suppresses the emission of bremsstrahlung radiation.

\subsection{Helicity-flip meson exchange estimations for $Q^2 \lesssim 0.5$ GeV${}^2$}
Besides affecting conventional radiative corrections, which are represented by the QED diagrams in Fig. 1(a)-(k), lepton mass effects in precision measurements of elastic $l^\pm p$ scattering with $\varepsilon_1 \lesssim m$ are expected to play a decisive role in $t$ channel meson exchanges that are mediated by the two-photon coupling of the meson (see, e.g., Fig. 3). Such contributions are called helicity-flip transitions because of a direct proportionality to the mass of the lepton and appear to be charge-odd whenever their interference with the OPE amplitude (Fig. 1(a)) is considered. If not estimated properly for the scattering of non-ultra-relativistic leptons, e.g. in case of muon scattering in the MUSE experiment, helicity-flip transitions would lead to substantial theoretical uncertainties. In our study \cite{KoshchiiSigma2016}, we showed that in the kinematical region of MUSE the largest inelastic helicity-flip contribution is expected from the respective scalar $\sigma$ meson exchange in the $t$ channel. This contribution was computed to be at most $\sim$ 0.1\% for muons, and it appeared to be about three orders of magnitude larger than for electrons (see Fig. 4 for details). This supports an idea that at the given level of precision, one can safely neglect respective contributions in scattering of ulra-relativistic electrons/positrons. To perform our estimation, we parameterized the coupling of the meson to two virtual photons by making use of the vector meson dominance model, which is well justified at $Q^2 \lesssim 0.5$ GeV${}^2$, including the kinematic region of MUSE. The calculation was done in part analytically and in part numerically using the LoopTools software \cite{HahnLoopTools1999}. Besides estimations of the respective charge-odd correction, we provided a first estimation of the effective coupling of the meson to the lepton.
\begin{figure}[h]
 \centerline{\includegraphics[width=100pt]{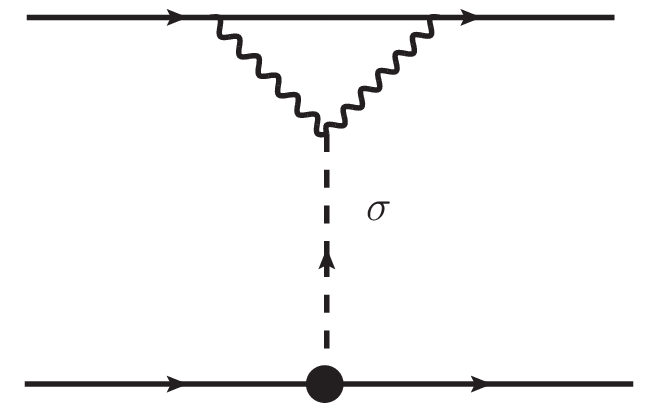}}
 \caption{$\sigma$ meson exchange in the $t$ channel.}
\end{figure}
\begin{figure}[h]
 \centerline{\includegraphics[width=330pt]{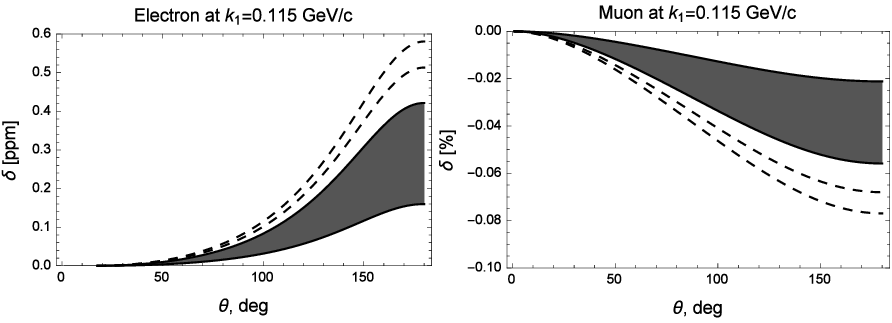}}
 \caption{Charge-dependent correction due to the interference between the OPE and $\sigma$ meson exchange in the $t$ channel amplitudes (shaded and transparent regions represent different models that were used to perform the calculation; see Ref. \cite{KoshchiiSigma2016} for details).}
\end{figure}
This estimation accounts only for the coupling of the virtual $\sigma$ meson to the transversely polarized photons in the vector meson dominance model.
\section{Transverse single-spin asymmetries in elastic electron-nucleon scattering}

Due to time-reversal invariance of the electromagnetic interaction, the transverse SSA observable $A_y^N$ in elastic lepton-nucleon scattering is zero in the Born approximation. Moreover, it can be shown \cite{RujulaSSA1971} that the leading contribution to this asymmetry is given by the following expression:
\begin{equation}\label{4}
    A_y^N = \frac{\mathrm{Im} \big[ T_{1 \gamma}^\dag \cdot \mathrm{Abs} (T_{2 \gamma})\big]}{ |T_{1 \gamma}|^2} \sim \int d^3 \vec{K}^* \ \mathrm{Im} \big( L^{\mu \alpha \beta} H_{\mu \alpha \beta}\big),
\end{equation}
where $L^{\mu \alpha \beta}$ and $H_{\mu \alpha \beta}$ are the leptonic and hadronic tensors, the integral is performed over the phase space of the intermediate lepton, and the proportionality coefficient is well-known. It is also convenient to split  $H_{\mu \alpha \beta}$ into two pieces: an elastic piece $H_{\mu \alpha \beta}^{el}$, which describes the intermediate nucleon state in the TPE loop (Fig. 1(j)), and an inelastic piece $H_{\mu \alpha \beta}^{in}$, which describes the intermediate state in the TPE loop that is not given by the nucleon. The greatest challenge in calculations of transverse asymmetries Eq. (\ref{4}) is the lack of knowledge about the hadronic inelastic tensor $H_{\mu \alpha \beta}^{in}$. This stems from the fact that in the most general case of scattering at non-forward angles the hadronic tensor consists of 18 gauge invariant tensor structures and 18 independent amplitudes \cite{Tarrach1975}, which we have little information about. As a result, a number of models and approaches exist that parameterize the respective tensor in different kinematical settings \cite{AfanasevBNSSA2004, PasquiniSSA2004, AfanasevTNSSA2002, BorisyukBNSSA2006, GorsteinBNSSA2008}. These models employ the knowledge on the behavior of certain amplitudes in various kinematical limits (see, e.g., Ref. \cite{Drechsel2003, Hagelstein2016}) and have been used to describe the measurements of beam-normal SSAs \cite{SampleBNSSA2001, A4Mass2005, G0BNSSA2007, HAPPEX2012}.

Recently, a first non-zero target-normal SSA measurement was performed at Jefferson Lab on a polarized ${}^3$He target \cite{ZhangSSA2015}. By using the fact that this target can be considered as an effective polarized neutron target, the authors of Ref. \cite{ZhangSSA2015} were able to extract a non-zero neutron-normal SSAs. The obtained asymmetries $A_y^n = -3.32 \pm 0.4 \pm 0.72 \%$, $A_y^n = -1.78 \pm 0.2 \pm 0.66 \%$, and $A_y^n =-1.38 \pm 0.14 \pm 0.24 \%$, which correspond to $\varepsilon_1 = 1.245$ GeV and $Q^2 = 0.127$ GeV${}^2$, $\varepsilon_1 = 2.425$ GeV and $Q^2 = 0.460$ GeV${}^2$, and $\varepsilon_1 = 3.605$ GeV and $Q^2 = 0.967$ GeV${}^2$, respectively, indicate that the inelastic TPE loop contribution is dominant in the considered kinematical region (our estimations of the elastic contribution are shown in Fig. 5). Currently, there exist no theoretical model that could be used for the description of neutron-normal SSAs at GeV beam energies and nearly forward scattering angles, including the results of Ref. \cite{ZhangSSA2015}. The development of respective unitarity-based approach is underway.

Measurements of transverse SSAs in elastic lepton-nucleon scattering provide an extremely valuable information on the imaginary part of the TPE amplitude and can be used to improve our understanding of the structure of the hadron, thus contributing to advances in theory. Moreover, since transverse SSAs are expected to be of an opposite sign for beams of positively and negatively charged leptons, future experimental data on such asymmetries would lead to studies of multi-photon exchange physics beyond TPE.
\begin{figure}[h]
 \centerline{\includegraphics[width=250pt]{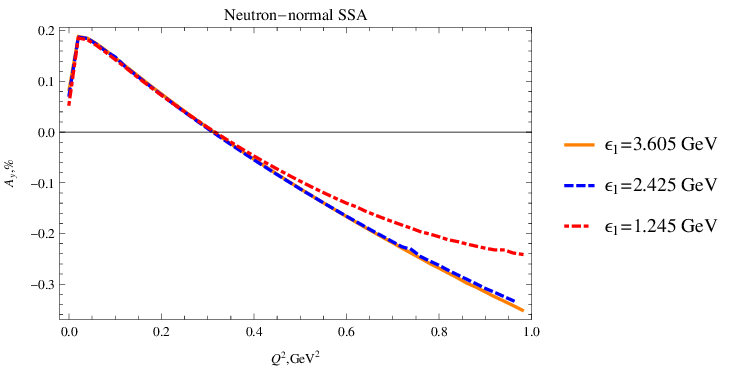}}
 \caption{Intermediate elastic contribution to the neutron-normal SSA.}
\end{figure}
\section{ACKNOWLEDGMENTS}
O.K. is grateful to the organizers of the workshop for the financial support during his stay at JLab. This work was supported in part by the Gus Weiss Endowment of The George Washington University and in part by a JSA/JLab Graduate Fellowship Award 2016/2017.


\nocite{*}
\bibliographystyle{aipnum-cp}%
\bibliography{sample}%

\end{document}